\documentclass[RNAAS]{aastex62}
\accepted{2018/05/02}
\received{2018/05/01}
\submitjournal{RNAAS}
\usepackage{graphicx}
\usepackage{natbib}
\usepackage{latexsym}
\usepackage{amssymb}
\usepackage{longtable}
\usepackage{amsmath}
\usepackage{url}
\begin{document}

\title{Transit time derivation for hot planet bow-shocks}

\author{P. Wilson Cauley}
\affiliation{Arizona State University\\
School of Earth and Space Exploration, Tempe, AZ 85287}

\author{Evgenya L. Shkolnik}
\affiliation{Arizona State University\\
School of Earth and Space Exploration, Tempe, AZ 85287}

\author{Joe Llama}
\affiliation{Lowell Observatory\\
Flagstaff, AZ 86001}

\correspondingauthor{P. Wilson Cauley}
\email{pwcauley@gmail.com}


\section{}

Measurements of exoplanet magnetic fields remain elusive, although detections
have now been confirmed in a small number of T-dwarfs
\citep{route12,route13,kao16}. While radio observations of auroral 
emission continue to be pursued, the possibility of
indirectly measuring planetary magnetic fields via UV and optical observations
of bow-shocks around hot planets remains a plausible path forward
\citep{vidotto10,llama11,benjaffel,cauley15}. The upcoming Colorado
Ultraviolet Transit Experiment (CUTE) aims, in part, to find such
signals\footnote{http://lasp.colorado.edu/home/cute/}. Recent work has also suggested
that estimates of hot Jupiter magnetic fields must be revised upward to take
into account the extra heat being deposited into their interiors
\citep{yadav17}. If these field strength estimates are accurate, many known
hot Jupiter systems should exhibit pre-transit interactions between the stellar
wind and the planetary magnetosphere. 

To aid in planning observations of signatures due to transiting bow-shocks, we
derive an estimate of the contact time of the bow-shock nose with the stellar
disk. This will allow telescope resources to be allocated more efficiently when
searching for such signatures.  The formula is a generalization of Eq. 19 from
\citet{vidotto11b}; the final form does not rely on a sky-projected value of
$r_\text{m}$.  We only consider leading shocks \citep[see][for
details]{vidotto11a}. Our result is only approximate: if the density along the
bow is sufficient, a portion of the bow away from the nose may cause a transit
signal before the nose reaches the stellar disk. For reasonable bow geometries
and densities, however, this should only amount to a difference of $\approx \pm
10$ minutes, which we consider negligible.  A circular orbit is assumed since
there is no closed-form for the distance along an arc of an ellipse. However,
planets with $e \lesssim 0.1$, which encompasses the majority of transiting
giant planets, should not show significant deviations from the derived
pre-transit time for circular orbits.

\begin{figure*}[b!]
   \centering
   \includegraphics[scale=.75,clip,trim=20mm 35mm 5mm 45mm,angle=0]{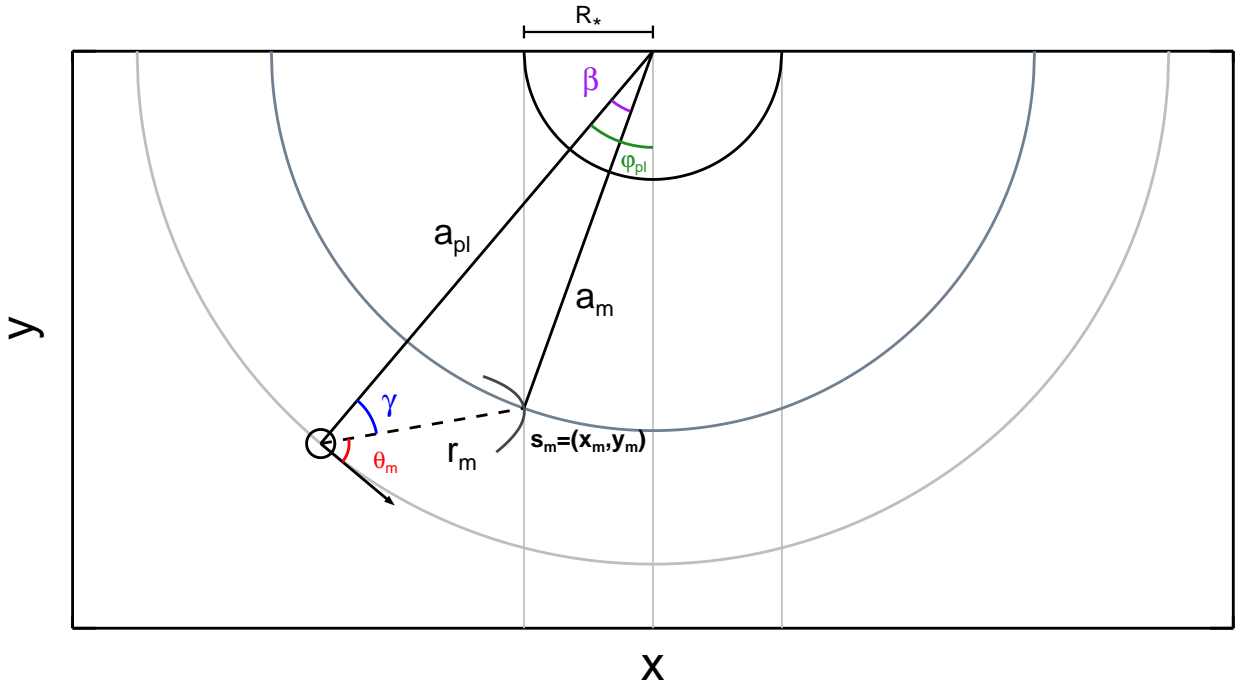}
   \figcaption{Bow-shock geometry.
\label{fig:fig1}}
\end{figure*}

\autoref{fig:fig1} shows the geometry of the transiting bow-shock. The planet
is represented by the small black circle and the bow is shown as a parabola.
The nose of the bow is at $\textbf{$s_\text{m}$} = (x_\text{m},y_\text{m})$
which is a distance $r_\text{m}$ away from the center of the planet. The bow
makes an angle $\theta_\text{m}$ with the tangent to the planet's orbit. This
angle is determined by the relative velocity of the planet and the stellar
wind \citep[see][]{vidotto11a}. The angle $\gamma = 90^\circ -
\theta_\text{m}$ is marked in blue and the angle $\beta$, the angle between the
star-planet line and star-bow nose line, in purple. The angle
$\phi_{\text{pl}}$, the planet's angular distance from
mid-transit, is shown in green. The distances $a_{\text{pl}}$ and $a_\text{m}$
represent the semi-major axis of the planet's orbit and the bow nose orbit,
respectively.

The bow nose transits when $x_\text{m} = -\sqrt{R_*^2 - b^2}$ where $b$ is the
planet's transit impact parameter in units of $R_*$ ($b=0$ in
\autoref{fig:fig1}. The position $x_\text{m}$ is given by

\begin{equation}\label{eq:eq1}
x_\text{m} = -a_\text{m} \sin(\phi_{\text{pl}} - \beta)
\end{equation} 

\noindent which, for the transit condition to be met, is

\begin{equation}\label{eq:eq2}
-\sqrt{R_*^2 - b^2} = -a_\text{m} \sin(\phi_{\text{pl}} - \beta).
\end{equation}

Using the law of cosines we can calculate $a_m$:

\begin{equation}\label{eq:am}
a_\text{m} = \sqrt{ a_{\text{pl}}^2 + r_\text{m}^2 - 2 a_{\text{pl}} r_\text{m} \cos\gamma } = \sqrt{ a_{\text{pl}}^2 + r_\text{m}^2 - 2 a_{\text{pl}} r_\text{m} \sin\theta_\text{m} }
\end{equation}

\noindent where we have replaced $\cos \gamma$ with $\sin \theta_\text{m}$. We can also express $\beta$ in terms of $r_\text{m}$, $a_\text{m}$, and $\theta_\text{m}$:

\begin{equation}\label{eq:beta}
\beta = \sin^{-1}\left(\frac{r_\text{m}}{a_\text{m}} \cos \theta_\text{m}\right).
\end{equation}

The planet's angle from mid-transit $\phi_\text{pl}$ can be expressed as the distance 
along the arc subtended by $\phi_\text{pl}$:

\begin{equation}\label{eq:phi}
\phi_\text{pl} = \frac{v_\text{orb} \Delta t}{a_\text{pl}}
\end{equation}

\noindent where $v_\text{orb}$ is the planet's Keplerian orbital velocity and $\Delta t$ is the time
from mid-transit for the planet.

We can now combine \autoref{eq:eq2}, \autoref{eq:beta}, and \autoref{eq:phi} to solve for
$\Delta t$ as a function of the planet and bow parameters:

\begin{equation}\label{eq:deltat}
\Delta t = \frac{v_\text{orb}}{a_\text{pl}} \left[ \sin^{-1} \left(\frac{\sqrt{R_*^2 - b^2}}{a_\text{m}}\right) + \sin^{-1}\left( \frac{r_\text{m}}{a_\text{m}} \cos \theta_\text{m} \right) \right]
\end{equation}

\noindent where $\Delta t$ is the beginning of the bow-shock transit relative
to the transit midpoint of the planet. We have left \autoref{eq:deltat} in
terms of $a_\text{m}$ for clarity, although $r_\text{m}$ is the more physically
relevant parameter \citep[see Eq. 9 of][]{llama13}. The value $a_\text{m}$ can
be calculated using \autoref{eq:am}.

\autoref{eq:deltat} can be used to estimate the expected transit time of the
bow-shock nose if approximations have been made for $r_\text{m}$ and
$\theta_\text{m}$. This may allow, for example, a half-night of telescope time
to be used to measure the bow-shock transit of a planet when $\Delta t$ is
small, rather than a full night. 

Most transiting hot Jupiter hosts do not have measured magnetic fields
\citep{fares13,mengel17}. This is important since the stellar field strength plays a
large role in determining $r_\text{m}$. Similarly, the stellar wind speed at
hot Jupiter orbital distances is poorly constrained for all stars. Thus
\autoref{eq:deltat} should be used to consider the range of plausible bow-shock
transit times given the range of reasonable values for the host star's magnetic
field and wind parameters. Finally, hot planets orbit through inhomogeneous
regions of the stellar wind which may result in $\Delta t$ changing
as a function of time as the star's activity level
varies \citep{llama13}. This should be taken into account when considering
values of $\Delta t$. 

\bigskip


\end{document}